\documentclass{kasper}
\usepackage{bm}
\usepackage[bottom]{footmisc}

\numberwithin{equation}{section}

\newcommand{\teight}[1]{t_8^{(#1)}}
\newcommand{\norder}[2]{:\!#1\!:\!#2\,}
\DeclareMathOperator{\tr}{tr}
\def\thbar{\theta\mathcal{C}}
\newcommand{\bPsi}{\overline\Psi}

\newdimen\tableauside\tableauside=1ex   
\newdimen\tableaurule\tableaurule=.32pt   
\newdimen\tableaustep
\def\phantomhrule#1{\hbox{\vbox to0pt{\hrule height\tableaurule width#1\vss}}}
\def\phantomvrule#1{\vbox{\hbox to0pt{\vrule width\tableaurule height#1\hss}}}
\def\sqr{\vbox{%
  \phantomhrule\tableaustep
  \hbox{\phantomvrule\tableaustep\kern\tableaustep\phantomvrule\tableaustep}%
  \hbox{\vbox{\phantomhrule\tableauside}\kern-\tableaurule}}}
\def\squares#1{\hbox{\count0=#1\noindent\loop\sqr
  \advance\count0 by-1 \ifnum\count0>0\repeat}}
\def\tableau#1{\vcenter{\offinterlineskip
  \tableaustep=\tableauside\advance\tableaustep by-\tableaurule
  \kern\normallineskip\hbox
    {\kern\normallineskip\vbox
      {\gettableau#1 0 }%
     \kern\normallineskip\kern\tableaurule}%
  \kern\normallineskip\kern\tableaurule}}
\def\gettableau#1 {\ifnum#1=0\let\next=\null\else
  \squares{#1}\let\next=\gettableau\fi\next}

\newcommand{\tabstack}[2]{\genfrac{}{}{0pt}{}{\vphantom{\tableau{1 1 1
1 1}}{#1}}{\mathbf{#2}}}
\newcommand{\oplusr}{\raise2ex\hbox{$\oplus$}}

\begin{document}
\preprintnumber{CERN-TH/2001-358\\
SPHT-T01/138\\
hep-th/0112157}
\title{Chiral splitting and world-sheet gravitinos in higher-derivative
string amplitudes}
\author{Kasper Peeters$^{1}$, Pierre Vanhove$^{1,2}$ and Anders Westerberg$^{1}$}
\address{1}{CERN\\
TH-division\\
1211 Geneva 23\\
Switzerland}
\address{2}{SPhT\\
Orme des Merisiers\\
CEA/Saclay\\
91191 Gif-sur-Yvette Cedex\\
France}
\date{December 18, 2001}
\onleave{2}{On leave of absence from SPhT, Saclay, France.}
\email{kasper.peeters, pierre.vanhove, anders.westerberg@cern.ch}
\maketitle

\begin{abstract}
  We report on progress made in the construction of higher-derivative
  superinvariants for type-II theories in ten dimensions. The string
  amplitude calculations required for this analysis exhibit
  interesting features which have received little attention in the
  literature so far. We discuss two examples from a forthcoming
  publication: the construction of the $(H_{\text{NS}})^2 R^3$ terms and the
  fermionic completion of the $\epsilon\epsilon R^4$ terms. We show
  that a correct answer requires very careful treatment of the chiral
  splitting theorem, implies unexpected new relations between
  fermionic correlators, and most interestingly, necessitates the use
  of world-sheet gravitino zero modes in the string vertex
  operators. In addition we compare the relation of our results to 
  the predictions of the linear scalar superfield of the type-IIB
  theory.
\end{abstract}
\maketoc
\begin{sectionunit}
\title{Introduction}
\maketitle

In a recent paper~\cite{Peeters:2000qj} we have initiated an
extensive project to determine the supersymmetric completion of the
higher-derivative expansion of string effective actions in ten
dimensions. Although several terms in these derivative expansions are
known, they concern almost exclusively the pure graviton sector,
except for a few isolated terms related to anomalies and a number of
terms conjectured from duality symmetries. Apart from the obvious
interest in extending our systematic results to include also the
Neveu-Schwarz~(NS-NS) and Ramond~(RR) gauge fields, we have argued
in~\cite{Peeters:2000qj} that there are several reasons why the
fermion bilinears are interesting as well. We have shown how, using a
combination of string scattering amplitudes and component
supersymmetry, one is able to determine these terms. The reader is
referred to the introduction of~\cite{Peeters:2000qj} for further
historical details and motivation.\footnote{Our previous paper also dealt with 
an extension of the ten-dimensional results to eleven dimensions, but we will 
not discuss this issue in the present letter.}

In the process of computing the string amplitudes which are required
for the full construction of the type-II invariants at order
$(\alpha')^3$, we realised that there are several interesting
subtleties which deserve additional attention. The aim of the present
letter is therefore to highlight these issues, separated from the
(necessarily rather technical) bulk of a future publication in which
the full invariants will be constructed. As we will show, our
calculations (and the various consistency checks they have to pass)
are an excellent way to probe some ill-appreciated features of the
formalism for (one-loop) string amplitudes in the RNS formalism. In
particular, we will highlight the role of left/right-mixing
contractions both in the amplitudes and in the picture-changing
procedure (particularly relevant in the odd spin-structure sector),
discuss in some detail the machinery necessary for the evaluation of
fermionic correlators, and exhibit the role of world-sheet gravitino
zero modes appearing in vertex operators.

Let us start with a short reminder of the present knowledge of
higher-derivative invariants in ten dimensions. The eight-derivative
terms of the superstring effective actions in ten dimensions are
constructed from two separate bosonic building blocks, which at linear
order in the NS-NS gauge field are given by
\begin{equation}
\label{e:twoinvariants}
\begin{aligned}
I_X &= t_8t_8 W^4 + \tfrac{1}{2}\varepsilon_{10} t_8 B W^4 \, ,\\[1ex]
I_Z &= -\varepsilon_{10}\varepsilon_{10} W^4 + 4\, \varepsilon_{10} t_8 B
W^4 \, .
\end{aligned}
\end{equation}
Here, $W$ denotes the Weyl tensor and $B$ the NS-NS
two-form gauge field. The tensors~$t_8$ and~$\varepsilon_{10}$ soak up
the free indices. These two building blocks are separately invariant
under a subset of the supersymmetry transformation rules, namely those
that lead to variations which are completely independent of any of the
gauge fields of the theory. This limited supersymmetry analysis does
not predict the relative coefficient between the two building blocks;
whether or not full supersymmetry fixes this coefficient remains an
open question. The relative normalisation has, however, been determined 
in other ways, and it turns out that its value in the various string 
theories is such that
\begin{equation}
\label{e:string_invariants}
\begin{aligned}
{\cal L}_{\text{heterotic}}\Big|_{(\alpha')^3} &= 
 e^{-2\phi^H} \big( I_X - \tfrac{1}{8} I_Z\big) 
+ I_X \, ,\\[1ex]
{\cal L}_{\text{IIA}}\Big|_{(\alpha')^3} &= e^{-2\phi^A} \big( I_X - \tfrac{1}{8} I_Z \big)
+ \big( I_X + \tfrac{1}{8} I_Z \big)\, ,\\[1ex]
{\cal L}_{\text{IIB}}\Big|_{(\alpha')^3} &=
 f(\Omega,\bar\Omega) \big(  I_X-\tfrac{1}{8}I_Z \big) \,.
\end{aligned}
\end{equation}
(Here $\Omega$ is the complexified coupling constant, and we have ignored
the Yang-Mills part of the heterotic effective action for simplicity). As a
result, the coefficient of the $B\wedge t_8 W^4$ term relative to the
$t_8 t_8 W^4$ term is zero for the type-IIB theory, while in the
type-IIA theory it is twice the coefficient of the heterotic theory.

In our previous paper, we have analysed in detail the fermionic
completion of the $I_X$ invariant and its associated supersymmetry
algebra.  In the present letter we discuss two specific calculations
that are needed in order to extend this result to other terms in the 
effective actions. The first one, discussed in section~\ref{s:iz}, deals
with the fermionic completion of the $I_Z$ invariant. Whereas the sign
flip of the $B\wedge t_8 W^4$ term between the two invariants
in~\eqn{e:twoinvariants} is simple to understand in terms of a GSO
projection sign, the related sign flip for the associated fermion
bilinears is very much hidden. This example is particularly relevant
in order to understand modifications to the supersymmetry algebra,
which was our original motivation for~\cite{Peeters:2000qj}.  

The second example concerns bosonic terms, namely the $(H_{\text{NS}})^2W^3$
terms of the type-II theories; these terms are discussed in
section~\ref{s:HHRRR}.  They have recently attracted attention in
their role as string-theory corrections to supergravity solutions
appearing in the AdS/CFT correspondence and non-conformal
gauge-theory/gravity dualities (see~\dcite{Frolov:2001xr} and
references therein).  The existing results on the $(H_{\text{NS}})^2
W^3$ terms are rather incomplete, mainly due to the fact that they
were obtained in the light-cone gauge. Moreover, a comparison of these
terms with the prediction of the linearised superfield construction
has not appeared so far. We exhibit a covariant calculation of these
terms and discuss how they might arise from the linearised superfield.

In both of these calculations, a precise understanding of the string
subtleties mentioned at the beginning of this introduction turns out to
be crucial in order to arrive at the correct result. The examples
discussed in the present letter are typical of the problems that arise
in the full construction of the type-II invariants, which will appear
elsewhere.
\end{sectionunit}

\begin{sectionunit}
\title{Picture changing, zero modes and contact terms}
\maketitle
\label{s:pco}

Chiral splitting---or the separation of string amplitudes in a
``left-moving, left-handed'' and a ``right-moving, right-handed''
factor---is often implicitly used in string calculations. As explained
by~\dcite{D'Hoker:1989ai}, the separation into two completely
independent sectors is, however, a subtle issue. On-shell conditions,
which constrain the fields to be either holomorphic or
anti-holomorphic functions of the complex world-sheet coordinate, can
obviously not be used inside the path integral. However, this is not
the only source of problems. For instance, the Green function of the
bosons on a torus implies a left/right-mixing two-point function
\begin{equation}
\label{e:lrcontract}
\big\langle \partial X(z) \bar\partial X(w) \big\rangle =
\alpha'\pi \Big( \delta^{(2)}(z-w)
- \frac{1}{\tau_2} \Big)\, ,
\end{equation}
receiving contributions both from a contact interaction and from zero modes. 
In addition the action contains a four-fermi term which does not respect the 
separation of Weyl components.

In practice, however, few of these things seem to matter. The reason
for this is reflected in the chiral splitting theorem
of~\dcite{D'Hoker:1989ai}. This theorem implies that there exist
\emph{effective} rules that summarise the consequences of the coupling
of both sectors. In many cases the implication is simply that
amplitudes can be obtained as the product of two chirally split
factors. However, this is definitely \emph{not} what happens in the
most general case. In the present section we would like to recall
these subtleties of the chiral splitting theorem and formulate them in
such a way that they can be readily applied to string amplitude
calculations. 

We start with the world-sheet action for the superstring. While the
graviton can always be gauge fixed to the flat metric, one has to be
careful with the gravitino; we therefore keep this field present. The
action for the \mbox{$(1,1)$-supersymmetric} string in a flat background is
then given by
\begin{multline}
S_{X,\Psi,F} =
\int\!{\rm d}^2z\, \Big(\partial X^\mu \bar\partial X_\mu 
- \Psi^\mu \bar\partial\Psi_\mu - \tilde\Psi^\mu \partial \tilde\Psi_\mu 
+F^\mu F_\mu \\[1ex]
+ \chi_- \bar\partial X^\mu \tilde\Psi_\mu 
+ \tilde\chi_+ \partial X^\mu \Psi_\mu 
+ \tfrac{1}{2}\Psi\tilde\Psi\,\chi_-\tilde\chi_+ \Big)\, .
\end{multline}
where $\Psi$ and $\chi$ denote the left-handed component of the
world-sheet fermion and gauge-fixed gravitino respectively, and tildes
denote the opposite chirality components.  In addition there are of
course the ghost terms.  Our gauge choice in the odd spin-structure
sector will use a delta-function localized gravitino slice,
\begin{equation}
\label{e:slice}
\tilde\chi_+(z) = \tilde\chi_+ \, \delta^{(2)}(z-y)\,, \qquad \chi_-(z) = \chi_-
\, \delta^{(2)}(z-y)\, ,
\end{equation}
and since we only work at genus one, there will be at most one such
zero mode for each chirality.  The one-loop amplitudes are independent
of the position of the slice~$y$ which will be chosen at convenience
when computing amplitudes. Note that we keep the auxiliary field
because a correct BRST treatment requires a symmetry algebra which
closes off shell.

Since the BRST charge contains the derivative of the scalar
fields~$X^\mu $, any verification of BRST invariance of vertex
operators also containing these fields is made more difficult by
the presence of the contraction~\eqn{e:lrcontract}. Therefore, it is
useful to start the analysis of the vertex operators in the ghost
picture for which they contain (apart from plane-wave exponentials)
only world-sheet spinors. These preferred graviton, two-form gauge
field and dilaton operators are given by
\begin{equation}
\label{e:Vcanonical}
V^{(-1,-1)}_\zeta = \int\!{\rm d}^2z\,\Big(\zeta_{\mu \nu}\, \Psi^\mu \tilde\Psi^\nu \,e^{-\phi-\tilde\phi}
e^{ik\cdot X}\Big)\,.
\end{equation}
For target-space fermions one has to take operators in the
$(-1/2,-1)$ picture to achieve a similar goal.
Using this class of vertex operators, all contractions of the
type~\eqn{e:lrcontract} are between picture-changing operators. These
are of two different kinds: those from the requirement of a fixed
total ghost charge, and those from the integration over odd
supermoduli. Although the effective contraction rules for the two are
the same, they arise in completely different ways, as we will now
show.

Let us first consider the second kind of insertion. Picture changers
of this type arise from an expansion of the action and subsequent Berezin
integration over the odd supermoduli.  Contractions of the
type~\eqn{e:lrcontract} between picture-changing terms can only occur
when both sectors are in the odd spin-structure sector. A generic
correlator takes the form
\begin{multline}
\label{e:generic_corr}
\Big\langle V_1(z_1)\cdots V_n(z_n)\Big\rangle = 
\int\! {\cal D}\chi{\cal D}\bar\chi
{\cal D}X {\cal D}\Psi\, \Big[V_1(z_1)\cdots V_n(z_n)\Big]\, \exp\big({-S[X,\Psi]}\big)\\[1ex]
\times e^{\phi+\tilde\phi}\,\bigg(
1 - \frac{1}{2\pi\alpha'}\int\!{\rm d}^2z\, \tilde\chi_+\Psi\partial X(z)
  - \frac{1}{2\pi\alpha'}\int\!{\rm d}^2z\, \chi_- \tilde \Psi \bar\partial X(z)
  + \frac{1}{4\pi\alpha'}\int\!{\rm d}^2z\, \tilde \chi_+\chi_-\Psi\tilde\Psi(z)\\[1ex]
  + \frac{1}{(2\pi\alpha')^2}\int\!{\rm d}^2w\!\! \int\!{\rm d}^2z\, 
     \tilde\chi_+\Psi\partial X(w) \chi_- \tilde \Psi \bar\partial X(z)
\bigg)\, ,
\end{multline}
where we have expanded in powers of the world-sheet gravitino (the
supersymmetry ghost~$e^{\phi}$ arises from the super-Beltrami
differentials). The Berezin integral has to be saturated with
gravitino zero modes, which can come either from the terms displayed
explicitly above, or from the vertex operators (more on the latter
option later). In the former case, this leads to insertion of
$\partial X\Psi$-type terms, which corresponds to an application of
the picture-changing operator.  Contractions between them which use
the $\delta^{(2)}(z-w)$ part of~\eqn{e:lrcontract}, in the fifth 
term above, are now immediately seen to cancel the fourth term. As a
consequence, one finds the effective rule that \emph{left/right
contractions between bosons of odd spin-structure picture changers
only involve the zero-mode part}.

Picture changers inserted by hand to balance the ghost charge behave
differently. One does not usually keep them explicit inside the path
integral, but instead works out the vertex operators in a new ghost
picture. This approach is, however, also plagued by subtleties arising from
the contraction~\eqn{e:lrcontract} and the presence of gravitino
modes.\footnote{The importance of a careful treatment of gravitino
modes was recently stressed again in the context of two-loop
calculations by~\ddcite{D'Hoker:2001nj}{D'Hoker:2001zp}.} We will
first re-examine this picture-changing procedure, after which we will
derive an effective rule for the contractions between the bosons that
arise from these picture changers and those that arise from the
odd-supermoduli integration.

The picture-changing procedure amounts to taking the BRST commutator with 
an insertion of the ghost zero mode $\xi$,
\begin{equation}
\label{e:pco_comm}
V^{(q+1,\tilde q)} = \frac{2}{\alpha'} \big[ Q^{\text{L}}_{\text{BRST}}, \xi\,V^{(q,\tilde q)} \big]\, ,
\end{equation}
with an analogous expression for the right-moving sector. In the presence of
world-sheet gravitino zero modes the formulation of an appropriate
$Q_{\text{BRST}}$ is troublesome, so we prefer to directly use the
BRST transformation rules. These can be found in~\dcite{Ohta:1986af}
and the $\gamma$-ghost dependent parts read in our conventions (we
will set~$\alpha'=2$ from now on)
\begin{equation}
\label{e:brstrules}
\begin{aligned}
\delta(\partial X^\mu ) &= \gamma \partial \Psi^\mu + \tilde \gamma \partial
\tilde\Psi^\mu \, , \\[1ex]
\delta(\Psi^\mu ) &= \gamma\Big(\partial X^\mu -
\tfrac{1}{2}\tilde\Psi^\mu \chi_- \Big) + \tilde\gamma F^\mu \, ,\quad
\smash{\begin{aligned}[b]
\delta(\bar\partial X^\mu ) &= \tilde \gamma\bar\partial\tilde\Psi^\mu +
\gamma\bar\partial\Psi^\mu \, ,\\[1ex]
\delta(\tilde \Psi^\mu ) &= \tilde\gamma\Big(\bar\partial X^\mu -
\tfrac{1}{2}\Psi^\mu \tilde\chi_+ \Big) + \gamma F^\mu \, .
\end{aligned}}\\[1ex]
\delta(F^\mu ) &= \tilde\gamma\Big( \bar\partial \Psi^\mu + \partial X^\mu 
\tilde\chi_+ - \tfrac{1}{2}\tilde\Psi^\mu \tilde\chi_+\chi_-\Big)
+ \gamma\Big( \partial\tilde\Psi^\mu + \bar\partial X^\mu \chi_- -
\tfrac{1}{2}\Psi^\mu \chi_-\tilde\chi_+\Big)\, ,\\[1ex]
\delta(e^{ikX}) &= i \big(\gamma\,k\cdot \Psi^\mu - \tilde
\gamma\,k\cdot \tilde\Psi^\mu \big) e^{ikX} \, .
\end{aligned}
\end{equation}
Note the terms proportional to the equations of motion on the first
line; these are absolutely crucial to arrive at gauge-invariant
results later. In the standard approach, when the BRST transformation
rules are obtained from an OPE with the BRST charge, these terms arise from
the left/right-mixing contraction~\eqn{e:lrcontract}.\footnote{The
left/right-mixing contraction~\eqn{e:lrcontract} arises for instance in
\begin{equation}
Q_{\text{BRST}}^{\text{L}}\,\xi\, V^{(-1,0)}_\zeta = \Big(\oint\!\frac{{\rm d}w}{2\pi i}\, e^{\phi-\chi}
\partial X^\mu \Psi_\mu (w) + \cdots\Big) \Big(\int\!{\rm d}^2z\, \zeta_{\mu \nu}
\Psi^\mu \bar\partial X^\nu e^{\chi-\phi}(z) + \cdots\Big)\, ,
\end{equation}
and after use of~\eqn{e:deltarep} this produces a term proportional to the
equation of motion. Similarly, the contraction~\eqn{e:FF} is used to
go from~\eqn{e:mone_zero} to~\eqn{e:V00B} below.}

Starting from the vertex operators~\eqn{e:Vcanonical} in their
canonical picture, we can now apply~\eqn{e:brstrules}, with the
symbolic rule that $\gamma\rightarrow e^{\phi}$. This corresponds to the
usual picture-changing prescription in the operator formalism, as
given in~\eqn{e:pco_comm}. The first increase of ghost number is only
made slightly non-standard by the presence of the gravitino
gauge-fixing functions. Continuing, to for instance $V^{(0,0)}$ or
beyond, the contraction~\eqn{e:lrcontract} is also going to play a
role. For the type-II operators, we find after the first step
\begin{align}
\label{e:mone_zero}
V^{(-1,0)}_\zeta &= \int\!{\rm d}^2z\,\zeta_{\mu \nu} \Big( \Psi^\mu \big( \bar\partial X^\nu - i
k\cdot \tilde\Psi\tilde\Psi^\nu + \tfrac{1}{2}\Psi^\nu\tilde\chi_+\big)
+ F^\mu \tilde\Psi^\nu \Big) e^{-\phi}e^{ik\cdot X}\,\, ,\\[1ex]
\label{e:mzero_one}
V^{(0,-1)}_\zeta &= \int\!{\rm d}^2z\,\zeta_{\mu \nu} \Big(\big(\partial X^\mu - i
k\cdot \Psi\Psi^\mu - \tfrac{1}{2}\tilde\Psi^\mu \chi_-\big)
\tilde\Psi^\nu + \Psi^\mu F^\nu \Big)
 e^{-\tilde\phi} e^{ik\cdot X}\,\, .
\end{align}
Some subtleties arise when the polarisation is a pure trace, but as we
will not need the dilaton operator here we refrain from discussing
them further (see also~\dcite{Terao:1987ux}).  Applying the
picture-changing operator once more, we now also encounter
contractions of the type~\eqn{e:lrcontract}. When the polarisation
tensor is taken to be that of the NS-NS two-form, the various terms
quadratic in fermions can be grouped by performing one partial
integration. One then obtains (using $H_{\mu \nu\rho}=3\,ik_{[\mu }
B_{\nu\rho]}$)
\begin{multline}
\label{e:V00B}
V^{(0,0)}_B = \int\!{\rm d}^2 z \, B_{\mu \nu} \Big(\partial
X^\mu \bar\partial X^\nu - k\cdot \Psi\Psi^\mu \,
k\cdot \tilde\Psi\tilde\Psi^\nu
+\tfrac{1}{2}\Psi^\mu \tilde\Psi^\nu\,\chi_-\tilde\chi_+\Big)\,e^{ik_\rho X^\rho}
\\[1ex]
-\frac{1}{6}\int\!{\rm d}^2z\, H_{\mu \nu\rho}\Big(
\Psi^\mu \Psi^\nu\Psi^\rho\tilde\chi_+ - \tilde\Psi^\mu \tilde\Psi^\nu
\tilde\Psi^\rho\chi_- 
\Big)\,e^{ik_\rho X^\rho}\\[1ex]
- \frac{1}{2}\int\!{\rm d}^2z\,
 H_{\mu \nu\rho}\Big( \Psi^\mu \Psi^\nu\bar\partial X^\rho 
- \tilde\Psi^\mu \tilde\Psi^\nu\partial X^\rho - 2 \Psi^\mu 
\tilde\Psi^\nu F^\rho \Big)
\, e^{ik_\rho X^\rho}\, .
\end{multline}
The last line contains the additional equation-of-motion terms that
were used by \dcite{Green:1988qu} in order to generate contact terms
(required to satisfy world-sheet supersymmetric Ward identities). Here
we see them arise naturally from the operator $V^{(-1,0)}$ after
picture changing.

Only with this full vertex operator will string amplitudes be gauge
invariant; we will discuss an explicit example in
section~\ref{s:HHRRR}. In particular, as we have already mentioned, it
is possible to saturate the Berezin integrals in~\eqn{e:generic_corr}
by using the gravitino modes that are present in the vertex
operator~\eqn{e:V00B} (note that after insertion of a delta-function
supported world-sheet gravitino, the second line of the above
expression will no longer contain a \mbox{$z$-integral}, but instead
will be localised at the position of the gravitino. Nevertheless, the
amplitude will be independent of the choice of the gravitino slice).
The last line of~\eqn{e:V00B} was also used in this gauge-invariant
form by~\dcite{Gutperle:1997gy}, although there the term vanishing
on-shell that is required to complete it was added by hand, not
derived from picture changing.

We should mention that the vertex operators in the various pictures,
including the gravitino terms and the equation-of-motion terms, also 
can be obtained directly from superspace, thereby avoiding the
picture-changing operation altogether. With the condition
$\Gamma^m\chi_m=0$ on the gravitino, the necessary ingredients are the
lowest-order supervielbein determinant, $E = e$, as well as the
supercovariant derivatives of the superfield $\Phi^\mu=X^\mu +\theta
\Psi^\mu + \bar\theta\tilde\Psi^\mu + \bar\theta\theta F^\mu $:
\begin{equation}
\begin{aligned}
D_- \Phi^\mu &= \tilde\Psi^\mu + \theta F^\mu + \bar\theta( \bar\partial X^\mu + \tfrac{1}{2} \tilde\chi_+
\Psi^\mu ) + \theta\bar\theta( -\bar\partial \Psi^\mu -\tfrac{1}{2}\tilde\chi_+
\partial X^\mu - \tfrac{1}{4}\tilde\chi_+ \chi_- \tilde\Psi^\mu )\, ,\\[1ex]
D_+ \Phi^\mu &= \Psi^\mu + \theta( \partial X^\mu + \tfrac{1}{2} \chi_-
\tilde\Psi^\mu ) - \bar\theta F^\mu + \theta\bar\theta( -\partial \tilde\Psi^\mu -\tfrac{1}{2}\chi_-
\bar\partial X^\mu - \tfrac{1}{4}\chi_- \tilde\chi_+ \Psi^\mu )\, .
\end{aligned}
\end{equation}
The vertex operators in the various pictures now follow from
$\tilde V = E\,D_-\Phi D_+\Phi \exp(k\Phi)$ and
\begin{equation}
V^{(-1,-1)} = \tilde V \Big\vert_{\substack{\theta=0\\[.5ex]\bar\theta=0}}\, ,\quad
V^{(0,-1)} = \int\!{\rm d}\theta\, \tilde V \Big\vert_{\substack{\theta=0\\[.5ex]\bar\theta=0}}\, ,\quad
V^{(0,0)} = \int\!{\rm d}\theta{\rm d}\bar\theta\, \tilde V \Big\vert_{\substack{\theta=0\\[.5ex]\bar\theta=0}}\, .
\end{equation}
This procedure was followed by~\dcite{D'Hoker:1987bh}, but the new
terms have to our knowledge not yet been used in actual amplitude
calculations.  It is important to understand that the superfield
approach only applies in the case of the NS-NS~vertex operators,
since it is only for these operators that a superspace expression
is available. The spacetime-gravitino vertex operator, for instance, is
expected to receive world-sheet gravitino terms at higher ghost
picture as well, but this can only be derived using the
picture-changing procedure based on the transformation
rules~\eqn{e:brstrules}. The same thing holds true for RR~gauge-field vertex operators.

Let us now return to the derivation of an effective rule for the boson
contractions between normal picture changers and those arising from
the integration over odd supermoduli. This proceeds simply by
inserting $(-1,0)$ and $(0,0)$ operators into the path
integral~\eqn{e:generic_corr}. For $(-1,0)$~operators the contraction
of the boson in~\eqn{e:mone_zero} with the second term in brackets
in~\eqn{e:generic_corr} leads to
$\zeta_{\mu \nu}\Psi^\mu \Psi^\nu\chi_-$, which is precisely cancelled
by the presence of the third term in the vertex operator. Again, the 
zero-mode part of the bosonic contraction is left. Similarly, for
$(0,0)$~operators, such cancellations occur. We have therefore deduced
the second effective rule, namely that \emph{left/right contractions
between bosons of odd spin-structure picture-changing operators and
bosons of vertex operators again only involve the zero-mode part}.

Finally, the amplitudes exhibit contractions between bosons of two
vertex operators. These contractions are not related to any special
cancellation mechanism, and therefore \emph{left/right contractions
between bosons of vertex operators involve both the contact term and
the zero-mode part}. However, the effective rule is different, due to
the fact that the $\delta$-function part of the contraction is
cancelled (for appropriate analytic continuation in the momenta) by
zeroes arising from the contraction of plane-wave exponentials, as
in~\eqn{e:expexp}. Again, only the zero-mode part
of~\eqn{e:lrcontract} survives.

\end{sectionunit}

\begin{sectionunit}
\title{$R^4$ invariants in IIA and IIB supergravity}
\maketitle
\label{s:string_amplitudes}
\begin{sectionunit}
\title{Gravitino bilinears in the $\epsilon\epsilon R^4$ invariant}
\maketitle
\label{s:iz}

Our first example illustrating the above mentioned string-amplitude
subtleties involves the computation of terms which are needed for the
fermionic completion of the~$I_Z$ invariant discussed in the
introduction. One of the most intriguing aspects of the
superinvariants is how terms with very different tensorial structures
talk to each other under supersymmetry variations.  For instance, an
$\epsilon_{10}$ tensor can in this way (via an intermediate Hodge
dualisation) be related to a $t_8$ structure.  In the present section
we will see a similar phenomenon, namely the appearance of a~$t_8$
tensor from a correlator with more than ten world-sheet fermions in
the odd spin-structure sector.

We will here focus on a particular fermionic bilinear in the action,
namely the one needed to cancel terms that arise from
variation under supersymmetry of the $B$-field in the anomaly term.
This variation produces (among other terms) a fermionic contribution
\begin{equation}
\delta_\epsilon (B\wedge t_8 W^4) \rightarrow \teight{s}\, (\bar\psi_m \Gamma^{mr_1\cdots r_8}\epsilon) 
W_{r_1r_2 s_1s_2} W_{r_3r_4 s_3s_4} W_{r_5r_6 s_5s_6} W_{r_7 r_8 s_7 s_8} \, ,
\end{equation}
which can only be cancelled by adding a fermion bilinear to the action. 
For the heterotic string the required term is
\begin{equation}
\label{e:gamma7}
S^{\text{heterotic}}_{\bar\psi\Gamma^{[7]}\psi_2 W^3} = \int\!{\rm d}^{10}x\,e\,
\teight{s}\, (\bar\psi_m \Gamma^{m r_1\cdots r_6} \psi_{s_7s_8})
W_{r_1r_2 s_1s_2} W_{r_3r_4 s_3s_4} W_{r_5r_6 s_5s_6}\, ,
\end{equation}
with the overall coefficient determined by the supersymmetry
transformation rules.  In~\cite{Peeters:2000qj} we have discussed the
origin of this fermion bilinear for the heterotic theory. Just like the
bosonic anomaly term, it arises from an amplitude in the odd
spin-structure sector.

In the type-II theories the situation is more complicated. There we
have both an odd/even and an even/odd spin-structure sector, with a
relative minus sign between the two in the type-IIB case. Therefore, the
bosonic~$B\wedge t_8 W^4$ term receives two contributions which add up
for the type-IIA theory while they cancel for the type-IIB theory
(see~\dcite{vafa7}). This is again reflected in the sign difference
for the one-loop invariants appearing
in~\eqn{e:string_invariants}. Because of the supersymmetry argument
sketched above, there should be a similar addition/cancellation
mechanism for the associated fermion bilinears, now given by
\begin{equation}
\label{e:gamma7II}
S^{\text{type-II}}_{\bar\psi\Gamma^{[7]}\psi_2 W^3} = \int\!{\rm
d}^{10}x\,e\,\teight{s}\, (\bar\psi^I_m \Gamma^{m r_1\cdots r_6}
\psi^I_{s_7s_8}) W_{r_1r_2 s_1s_2} W_{r_3r_4 s_3s_4} W_{r_5r_6
s_5s_6}\, .
\end{equation}
Here $I=1,2$ labels the two Majorana-Weyl gravitinos. However, the
cancellation cannot simply be between the odd/even and even/odd
sectors. The reason is that these two sectors produce terms with
fermions of opposite spacetime chirality, which can never add up or 
cancel. Thus, a new mechanism is required, and we will show below
that it depends crucially on the left/right mixing discussed in the previous
section, as well as on some intricate identities between world-sheet
fermion correlators.

The only way in which terms of the form~\eqn{e:gamma7II} can be made
to add up or cancel for \emph{both} fermion types in the same way, is to
have them arise from the odd/odd spin-structure sector. Amplitudes
computed there differ by a minus sign (due to the GSO projection)
between the type-IIA and type-IIB theories. The surprising element is
that the odd/odd spin-structure sector in fact produces
the~$t_8$~tensor of~\eqn{e:gamma7II}, something that is not at all
obvious from the RNS fermionic correlator. This new odd/odd term which
we expect to contribute to the coefficient of~\eqn{e:gamma7II} is given 
by the following five-point amplitude:
\begin{multline}
\label{e:fullA5oo}
{\cal A}_5^{\text{odd/odd}}= \bigg\langle 
Y(w_1) \tilde Y(\tilde w_1)\,
\prod_{i=2}^4 \oint\!\frac{{\rm d}w_i}{2\pi i}\, Y(w_i)\, 
\prod_{j=2}^5 \oint\!\frac{{\rm d}\tilde w_j}{2\pi i}\, \tilde Y(\tilde w_j)\,\\[1ex]
\times \prod_{m=1}^2 \int\!{\rm d}^2 y_m\, V^{(-1/2,-1)}_{\psi_m}(y_m)\,
\prod_{n=1}^3 \int\!{\rm d}^2 z_n\, V^{(-1,-1)}_{g_n}(z_n)\bigg\rangle_{\text{odd/odd}}\, .
\end{multline}
We have written it in its most general form, with all the vertex
operators in their canonical pictures and the two odd spin-structure
picture-changing operators inserted at arbitrary points. 

Part of this amplitude has been computed previously
by~\dcite{Lin:1990xb}, namely the terms which do not involve a bosonic
zero-mode contraction of two odd spin-structure picture-changing
operators.  However, there is no~$\Gamma^{[7]}$ contribution from
these terms.  Indeed, there is in general no way in which the full
left/right separation can lead to a result which has a gravitino index
contracted with the gamma matrix, as in~\eqn{e:gamma7}.  We will
therefore immediately turn our attention to the case in which the two
picture-changing operators are contracted through their bosonic zero
modes.

In order to evaluate this part of the amplitude it is convenient to 
first eliminate some of the picture-changing contour integrals. Having
chosen the two picture changers which are to be contracted through
their bosonic zero modes, all the bosons of the remaining picture
changers are to be contracted with plane-wave exponentials (to produce
the $(0,0)$ ghost-picture operators and the associated effective
contraction rules which we discussed in section~\ref{s:pco}).
The odd/odd spin-structure contribution to the left/right-mixing part 
of the amplitude then reduces to the much simpler expression
\begin{equation}
\label{e:A5lroo}
\begin{aligned}
{\cal A}_5^{\text{odd/odd}}&\Big|_{\text{left/right}} = 
(\bar\psi_{m}^{(1)})_a (\psi_{np}^{(2)})_b\,
W^{(3)}_{r_1r_2s_1s_2}\,W^{(4)}_{r_3r_4s_3s_4}\,W^{(5)}_{r_5r_6s_5s_6}\\[1ex]
&\begin{aligned}
\times \bigg\langle 
  \Psi^l(w_1)\,\, &
  S^a(y_1)\, && \,\,\,\,\,S^b(y_2)\,  &&
  \norder{\Psi^{r_1}\Psi^{r_2}}{(z_1)}\, &&
  \norder{\Psi^{r_3}\Psi^{r_4}}{(z_2)}\, &&
  \norder{\Psi^{r_5}\Psi^{r_6}}{(z_3)}\\
  \tilde\Psi^l(\tilde w_1)\,\, &
  \tilde\Psi^{m}(y_1)\, &&
  \norder{\tilde\Psi^{n}\,\tilde\Psi^{p}}{(y_2)}\, &&
  \norder{\tilde\Psi^{s_1}\tilde\Psi^{s_2}}{(z_1)}\, &&
  \norder{\tilde\Psi^{s_3}\tilde\Psi^{s_4}}{(z_2)}\, &&
  \norder{\tilde\Psi^{s_5}\tilde\Psi^{s_6}}{(z_3)}
\bigg\rangle_{\text{odd}}
\end{aligned}\\[1ex]
&\times \bigg\langle
e^{\phi(w_1)}\,e^{-\phi(y_1)/2}\,e^{-\phi(y_2)/2}
\bigg\rangle_{\text{odd}}
\,
\bigg\langle
e^{\tilde\phi(\tilde w_1)}\,e^{-\tilde\phi(\tilde y_1)}
\bigg\rangle_{\text{odd}}
\, .
\end{aligned}
\end{equation}
Integration over the insertion points as in~\eqn{e:fullA5oo} is implicitly 
understood from now on. We have also replaced the linearised expressions
involving the graviton polarisations with the full Weyl
tensors. Bracketed superscripts on the polarisation tensors are used
to identify the five external states.

The above amplitude is to be added to
\begin{equation}
\label{e:A5lroe}
\begin{aligned}
{\cal A}_5^{\text{odd/even}}&\Big|_{\text{left/right}} = 
(\bar\psi^{(1)}_{m})_a (\psi^{(2)}_{np})_b\,
W^{(3)}_{r_1r_2s_1s_2}\,W^{(4)}_{r_3r_4s_3s_4}\,W^{(5)}_{r_5r_6s_5s_6}\\[1ex]
&\begin{aligned}
\times \; \bigg\langle &
  \Psi^l(w_1)\,
  S^a(y_1)\,&&\;\;\;\;\; S^b(y_2)\, &&
  \norder{\Psi^{r_1}\Psi^{r_2}}{(z_1)}\, &&
  \norder{\Psi^{r_3}\Psi^{r_4}}{(z_2)}\, &&
  \norder{\Psi^{r_5}\Psi^{r_6}}{(z_3)}\bigg\rangle_{\text{odd}}\\
\times \; \bigg\langle{} &
\;\;\;\;\frac{\eta^{lm}}{\bar w_1-\bar y_1} &&
  \norder{\tilde\Psi^{n}\,\tilde\Psi^{p}}{(y_2)}\,&&
  \norder{\tilde\Psi^{s_1}\tilde\Psi^{s_2}}{(z_1)}\,&&
  \norder{\tilde\Psi^{s_3}\tilde\Psi^{s_4}}{(z_2)}\,&&
  \norder{\tilde\Psi^{s_5}\tilde\Psi^{s_6}}{(z_3)}
\bigg\rangle_{\text{even}}\end{aligned}\\[1ex]
&\times \bigg\langle
e^{\phi(w_1)}\,e^{-\phi(y_1)/2}\,e^{-\phi(y_2)/2}
\bigg\rangle_{\text{odd}}
\,
\bigg\langle
e^{\tilde\phi(\tilde w_1)}\,e^{-\tilde\phi(\tilde y_1)}
\bigg\rangle_{\text{even}}
\, ,
\end{aligned}
\end{equation}
which arises by considering the even spin-structure sector for the
right-moving fermions.  Note that the difference with the odd/odd
amplitude~\eqn{e:fullA5oo} is minimal: the non-integrated
picture-changing operator on the right was dropped in exchange for an
integrated one. As we have already observed that the integrand
of~\eqn{e:A5lroo} is independent of the insertion point~$\tilde
w_1$,~it is clear that we need a different type of contraction in
order to obtain the odd/even amplitude. Instead of keeping the zero
modes on the right, we have now contracted one picture-changing
operator fermion with a fermion from the gravitino. Explicitly, this
means that we consider the $l$-$m$ contraction. This produces a pole
and therefore a non-vanishing result after integration over the
picture-changer insertion point. An epsilon tensor arises from the
left-handed sector. Having contracted these two fermions, the tensor
structure for the right-handed fermion sector is then fixed by
well-known symmetry arguments which we will not repeat.

The total amplitude can thus be written in a compact way as
\begin{equation}
\label{e:A5ooT}
{\cal A}_5\Big|_{\text{left/right}} = (\bar\psi^{(1)}_{m}\, T^{r_1\cdots r_6}{}_l \psi^{(2)}_{s_7s_8})\,
W^{(3)}_{r_1r_2s_1s_2}\,W^{(4)}_{r_3r_4s_3s_4}\,W^{(5)}_{r_5r_6s_5s_6} \,
\times \begin{cases}
\varepsilon_{(10)}^{lm s_1\cdots s_8} & \text{odd/odd}\, ,\\[1ex]
\eta^{lm} t_8^{s_1\cdots s_8} &\text{odd/even}\, ,
\end{cases}
\end{equation}
where---taking into account the symmetries imposed by the contracting
tensors---$T$ has the gamma-matrix expansion\footnote{In addition to
the complete symmetry in the exchange of $r$-index pairs and the
pair-wise antisymmetries, we also use the observation that any
contraction between $l$ and an $r$ index leads to a Weyl tensor
antisymmetrised in three indices and hence vanishes.}
\begin{equation}
\label{e:Texpansion}
\begin{aligned}
T^{r_1\cdots r_6\,l} &=
     A(y_i, z_i, w_i)\, \Gamma^{r_1\cdots r_6} \Gamma^l \\[1ex]
  &+ \tfrac{1}{2}\,C_1(y_i, z_i, w_i)\,\big(\eta^{r_1r_3}\eta^{r_2r_4}\Gamma^{r_5r_6}
      + \text{5 permutations} \big)\Gamma^l\\[1ex]
  &+ \tfrac{1}{2}\,C_2(y_i, z_i, w_i)\,\big(\eta^{r_1r_3}\eta^{r_2r_5}\Gamma^{r_6r_4}
      + \text{23 permutations} \big)\Gamma^l \,.
\end{aligned}
\end{equation}
For the odd/even amplitude we have to focus on the $\Gamma^{[7]}$
terms that arise from these expressions, while for the odd/odd
amplitude it is the $\Gamma^{[3]}$ terms that are relevant to obtain a
contribution to the effective action of the form~\eqn{e:gamma7II}. In 
the former case we simply have to compute the coefficient function~$A$.
In the latter case we first have to eliminate the epsilon tensor
in~\eqn{e:A5ooT} by dualising the gamma matrix onto which it is
contracted; this leads to
\begin{equation}
\label{e:dualgammal}
\varepsilon_{(10)}^{s_1\cdots s_8lm}\,\Gamma_l\, \psi_{s_7s_8}
  = \Gamma_{[9]}^{s_1\cdots s_8m}\,\psi_{s_7s_8}
  = \widetilde{\cal E}(\psi) - 42\, \Gamma_{[5]}^{[s_1\cdots s_5} \,
        \psi_{\phantom{[5]}}^{s_6m]}\, ,
\end{equation}
where $\widetilde{\cal E}(\psi)$ denotes the gravitino equation of
motion. Multiplying the right-hand side with the other gamma matrices
that appear in the decomposition of the $T$~tensor, we find
that the restriction of the five-point amplitude to $\Gamma^{[7]}$
terms in the odd/odd sector is independent of $A$,
\begin{multline}
{\cal A}_5^{\text{odd/odd}}\Big|_{\text{left/right}, \Gamma^{[7]}}  \\[1ex]
\begin{aligned}
{}&=  -6\,\bar\psi^{(1)}_m\Gamma^{mr_1\cdots r_6} \Big( C_1 \,
\tr\big(W^{(3)}_{r_1r_2} W^{(4)}_{r_3r_4}\big)\tr(W^{(5)}_{r_5r_6}\psi^{(2)}\big)
- 4\,C_2\,
\tr\big(W^{(3)}_{r_1r_2} W^{(4)}_{r_3r_4} W^{(5)}_{r_5r_6} \psi^{(2)}\big)\Big)\,,
\end{aligned}
\end{multline}
where permutation of the three graviton polarisation tensors is understood.
Provided we can show that the coefficient functions satisfy $C_1=C_2=C$,
the resulting amplitude will thus take the form
\begin{equation}
{\cal A}_5^{\text{odd/odd}}\Big|_{\text{left/right}, \Gamma^{[7]}}{} =
  C\,t_8^{s_1\cdots s_8} (\bar\psi^{(1)}_m\Gamma^{m r_1\cdots r_6}\psi^{(2)}_{s_7s_8})\,
  W^{(3)}_{r_1r_2s_1s_2}W^{(4)}_{r_3r_4s_3s_4}W^{(5)}_{r_5r_6s_5s_6}\, ,
\end{equation}
and can potentially cancel the odd/even part of the amplitude for the
type-IIB string.

In order to determine the precise expressions for $A$, $C_1$ and $C_2$,
we evaluate the amplitude for particular values of the $r$-, $a$- and
$b$-indices (for a combination which yields a non-zero element of the
gamma matrix under consideration, employing the helicity basis
of appendix~\ref{a:helicity}). These are listed, together with the
fermion combinations that appear in the resulting correlators, in
table~\ref{f:fermions}. 
\begin{table}[th]
\begin{small}\begin{equation*}
\begin{matrix} 
r_1=1    & r_2=3    & r_3=1,2   & r_4=3,4   & r_5=5    & r_6=8    & l=7      & a          & b       \\[2ex]
\Psi(z_1)& 0        & \bPsi(z_2)& 0         & 0        & 0        & 0        & S_+(y_1)   & S_-(y_2)  \\[1ex] 
0        & \Psi(z_1)& 0         & \bPsi(z_2)& 0        & 0        & 0        & S_+(y_1)   & S_-(y_2)  \\[1ex]
0        & 0        & 0         & 0         & \Psi(z_3)& 0        & 0        & S_-(y_1)   & S_-(y_2)  \\[1ex]
0        & 0        & 0         & 0         & 0        &\bPsi(z_3)&\Psi(w_1)& S_+(y_1)   & S_-(y_2)   \\[1ex]
0        & 0        & 0         & 0         & 0        & 0        & 0        & S_+(y_1)   & S_-(y_2)  \\[4ex] 
r_1=1    & r_2=3    & r_3=1     & r_4=8    & r_5=5     & r_6=3    & l=7      & a          & b       \\[2ex]
\Psi(z_1)& 0        & \bPsi(z_2)& 0        & 0         & 0        & 0        & S_+(y_1)   & S_-(y_2)  \\[1ex] 
0        & \Psi(z_1)& 0         & 0        & 0         &\bPsi(z_3)& 0        & S_+(y_1)   & S_-(y_2)  \\[1ex]
0        & 0        & 0         & 0        & \Psi(z_3) & 0        & 0        & S_-(y_1)   & S_-(y_2)  \\[1ex]
0        & 0        & 0         &\bPsi(z_2)& 0         & 0        &\Psi(w_1) & S_+(y_1)   & S_-(y_2)  \\[1ex]
0        & 0        & 0         & 0        & 0         & 0        & 0        & S_+(y_1)   & S_-(y_2)   
\end{matrix}
\end{equation*}
\end{small}
\vskip 1ex
\begin{equation*}
\begin{aligned}
(\Gamma^{57} \Gamma^8)_{ab} &= \sigma^2\otimes\sigma^1\otimes\mathbb{1}\otimes\sigma^1\otimes\sigma^1\, , \\[1ex]
(\Gamma^{1324578})_{ab} &=     \sigma^1\otimes \sigma^2\otimes\mathbb{1}\otimes\sigma^1\otimes\sigma^1\, .
\end{aligned}
\end{equation*}
\caption{The fermion distributions used for the computation of ${\cal
A}^{\text{odd/odd}}_5$ and ${\cal A}^{\text{odd/even}}_5$, together
with the gamma-matrix products that determine them. The upper table is
used to determine~$C_1$ and~$A$, while the bottom table shows the
correlator used for the computation of~$C_2$. The fermions are defined
in~\eqn{e:fermiondef}.}
\label{f:fermions}
\end{table}
The fermionic part of the amplitude can now be deduced using the
approach of \dcite{Atick:1987rs} or an equivalent construction based
on bosonisation.  Since the result is by construction independent of
the picture-changer insertion point, we can evaluate it for $w_1=z_1$
and find indeed that $C_1=C_2$.

Unfortunately, it is not so easy to see whether the coefficient
function~$A$ is related to~$C$ in such a way as required by supersymmetry. 
Although the distribution for the fermions is similar, 
in the sense that $(\Gamma^{1324578})_{ab}$ leads to the same type of
correlators as $(\Gamma^{57} \Gamma^8)_{ab}$, they are not
identical. It therefore seems that cancellation of the terms
in~\eqn{e:gamma7II} appears only after the insertion-point integrals
have been performed.\footnote{In addition, in order to arrive at the
  precise coefficient in the effective action, one should subtract a 
  five-point amplitude given by a tree-level graviton exchange graph 
  formed from the $R^4$ and $\bar\psi\Gamma^{[3]}\psi_{(2)}$ vertices.}

The fact that both odd/even and odd/odd amplitudes contribute to the
same term in the effective action will also be relevant for other
fermion bilinears. It is indeed a crucial observation in order to
construct the~$I_Z$ invariant as an effective action arising from
string amplitudes. More details about cancellation mechanisms
analogous to the one described here will appear in a forthcoming
publication.
\end{sectionunit}

\begin{sectionunit}
\title{Interactions involving the NS-NS two-form}
\maketitle
\label{s:HHRRR}
\begin{sectionunit}
\title{String amplitudes}
\maketitle

While the higher-derivative terms involving only gravitons have been
known now for quite some time, substantially less is known about terms
which involve NS-NS or RR gauge fields.\footnote{These terms are
  closely related to the eleven-dimensional terms which involve the
  four-form field strength. For all applications so far (see for
  instance~\dcite{Becker:2001pm}) the knowledge about only the anomaly
  term $C\wedge t_8 W^4$ was sufficient. A large-volume limit was
  taken, and in this limit all other terms involving the gauge field
  scale to zero faster than the anomaly term.}  In the present section
we will discuss the terms in the effective action which contain two
powers of the NS-NS three-form field strength and three powers of the
Weyl tensor.  We will not construct the full effective action here (we
ignore several subtraction problems), but focus instead on those
amplitudes which show that the subtleties discussed in
section~\ref{s:pco} are \emph{crucial} in order to arrive at the
correct answer.

The terms under consideration have been used in a recent paper
by~\dcite{Frolov:2001xr} in the context of gauge-theory/gravity duality. 
We would like to comment here in some detail on the
covariant calculation of several terms in the required amplitudes,
since they all originate from the picture-changing subtleties
discussed in section~\ref{s:pco}. As in the previous section we
restrict ourselves to a one-loop analysis.

In the odd/odd spin structure we find the result
\begin{multline}
{\cal A}_{\text{odd/odd}} = 
k^{(1)}_{r_7} B^{(1)}_{r_8 s_9} 
k^{(2)}_{s_7} B^{(2)}_{s_8 r_9} \,
W^{(3)}_{r_1r_2s_1s_2}
W^{(4)}_{r_3r_4s_3s_4}
W^{(5)}_{r_5r_6s_5s_6}\\[1ex]
\begin{aligned}
\times \Big\langle 
\partial X\cdot \Psi(w_1)\, & \norder{\Psi^{r_7} \Psi^{r_8}}{(z_1)}\, &&
\,\,\,\,\Psi^{r_9}(z_2)\, &&
\norder{\Psi^{r_1} \Psi^{r_2}}{(z_3)}\, &&
\norder{\Psi^{r_3} \Psi^{r_4}}{(z_4)}\, &&
\norder{\Psi^{r_5} \Psi^{r_6}}{(z_5)}\,\\
\bar \partial X\cdot \tilde\Psi(w_2)\, &
\,\,\,\,\tilde\Psi^{s_9}(z_1)\, &&
\norder{\tilde \Psi^{s_7} \tilde \Psi^{s_8}}{(z_2)}\, &&
\norder{\tilde \Psi^{s_1} \tilde \Psi^{s_2}}{(z_3)}\, &&
\norder{\tilde \Psi^{s_3} \tilde \Psi^{s_4}}{(z_4)}\, &&
\norder{\tilde \Psi^{s_5} \tilde \Psi^{s_6}}{(z_5)}
\Big\rangle\, ,
\end{aligned}
\end{multline}
where picture changers from the integration over the odd supermoduli
were inserted. Inserting the zero-mode term in~\eqn{e:lrcontract}
leads to a non-zero amplitude. In order to reproduce this amplitude
from an effective action, one has to take into account not just~$H^2
W^3$ terms, but also $(D H)^2 W^2$ terms, as the latter can produce
non-trivial five-point amplitudes as well. In particular, the above
amplitude can be reproduced from the field-theory action
\begin{equation}
\label{e:oddodd}
{\cal L}^{1}_{\text{odd/odd}} = \epsilon^{r_1\cdots r_9m} \epsilon^{s_1\cdots s_9}{}_m
W_{r_1r_2s_1s_2}W_{r_3r_4s_3s_4}W_{r_5r_6s_5s_6} \Big(
H_{r_7r_8s_9} H_{s_7s_8r_9} - \tfrac{1}{9}H_{r_7r_8r_9} H_{s_7s_8s_9}\Big)\, ,
\end{equation}
(which despite the many antisymmetrisations is not a total derivative)
or alternatively from
\begin{equation}
\label{e:GSanalogue}
{\cal L}^{2}_{\text{odd/odd}} = \epsilon^{r_1\cdots r_8mn}
\epsilon^{s_1\cdots s_8}{}_{mn} W_{r_1r_2s_1s_2}W_{r_3r_4s_3s_4}\, 
D_{s_5} H_{r_7 r_8 s_6}\, 
D_{r_5} H_{s_7 s_8 r_6}\, ,
\end{equation}
(in the latter case, the fifth graviton arises from a fluctuation of
one of the metrics contracting the two epsilon tensors).  At the level
of the five-point amplitude computed above, these contributions to the
effective action cannot be distinguished, though six-point amplitudes
will lift the ambiguity. The term~\eqn{e:GSanalogue} is similar to the
$t_8 t_8 (DH)^2 W^2$ term found by~\dcite{Gross:1987mw}, but note that
there is no contribution to four-point amplitudes
from~\eqn{e:GSanalogue}, which is why it has not been discussed in the
literature before.\footnote{As explained in e.g.~\dcite{Gross:1987mw},
  one also needs to subtract five-point functions obtained as
  tree-level graphs formed from lower-point vertices before a link
  with the effective action can be made.  Fortunately, in our case,
  the fact that these lower-order vertices come with two derivatives
  implies that one needs at least four three-point vertices in order
  to produce an eight-derivative amplitude. Such graphs have, however,
  at least six external lines.  This subtraction problem is therefore
  absent for the terms under consideration. Because our main focus is
  on the presence of~\eqn{e:oddodd}, and because the coefficient
  of~\eqn{e:GSanalogue} is anyhow undetermined, we will also not
  consider the subtraction problem associated with graphs formed from
  one vertex obtained from the~$R$ or~$H^2$ term and one vertex
  obtained from the~$t_8t_8(DH)^2 W^2$~term.}

In the odd/even sector (and similarly in the even/odd sector) there
are three different terms that contribute to the amplitude. Two of
them are standard in the sense that they only involve the 
left/right-mixing contraction~\eqn{e:lrcontract}. The third one is unusual
because it relies on the presence of world-sheet gravitino terms in
the $V^{(0,0)}_B$ vertex operator. Since the odd/even contribution
is parity odd, one expects it to be absent from the effective
action. The fact that the sum of the three terms indeed vanishes (for
both the type-IIA as well as the type-IIB theory) shows that the
world-sheet gravitino terms in vertex operators cannot be ignored.

Let us first discuss the two standard terms, one given by
\begin{multline}
{\cal A}^1_{\text{odd/even}} = \tfrac{3}{2}
k^{(1)}_{[r_7} B^{(1)}_{r_8 m]} 
k^{(2)}_{s_7} B^{(2)}_{r_9 s_8} \,
W^{(3)}_{r_1r_2 s_1 s_2}
W^{(4)}_{r_3r_4 s_3 s_4}
W^{(5)}_{r_5r_6 s_5 s_6}\\[1ex]
\begin{aligned}
\times \Big\langle 
\partial X\cdot \Psi(w_1)\, & \norder{\Psi^{r_7} \Psi^{r_8}}{(z_1)}\, &&
\,\,\,\Psi^{r_9}(z_2)\, &&
\norder{\Psi^{r_1} \Psi^{r_2}}{(z_3)}\, &&
\norder{\Psi^{r_3} \Psi^{r_4}}{(z_4)}\, &&
\norder{\Psi^{r_5} \Psi^{r_6}}{(z_5)}\,\\
{~} & \,\,\,\bar\partial X^m(z_1)\, &&
\norder{\tilde \Psi^{s_7} \tilde \Psi^{s_8}}{(z_2)}\, &&
\norder{\tilde \Psi^{s_1} \tilde \Psi^{s_2}}{(z_3)}\, &&
\norder{\tilde \Psi^{s_3} \tilde \Psi^{s_4}}{(z_4)}\, &&
\norder{\tilde \Psi^{s_5} \tilde \Psi^{s_6}}{(z_5)}
\Big\rangle\, ,
\end{aligned}
\end{multline}
and the other one obtained by using a different part of the 
gauge-field vertex operator:
\begin{multline}
{\cal A}^2_{\text{odd/even}} = 
k^{(1)}_{r_7} k^{(1)}_{s_7} B^{(1)}_{r_8 s_8} 
B^{(2)}_{r_9 m} \,
W^{(3)}_{r_1r_2 s_1 s_2}
W^{(4)}_{r_3r_4 s_3 s_4}
W^{(5)}_{r_5r_6 s_5 s_6}\\[1ex]
\begin{aligned}
\times \Big\langle 
\partial X\cdot \Psi(w_1)\, & \norder{\Psi^{r_7} \Psi^{r_8}}{(z_1)}\, &&
\,\,\,\Psi^{r_9}(z_2)\, &&
\norder{\Psi^{r_1} \Psi^{r_2}}{(z_3)}\, &&
\norder{\Psi^{r_3} \Psi^{r_4}}{(z_4)}\, &&
\norder{\Psi^{r_5} \Psi^{r_6}}{(z_5)}\,\\
{~} & \norder{\tilde \Psi^{s_7} \tilde \Psi^{s_8}}{(z_1)}\, &&
\,\,\,\bar\partial X^m(z_2)\, &&
\norder{\tilde \Psi^{s_1} \tilde \Psi^{s_2}}{(z_3)}\, &&
\norder{\tilde \Psi^{s_3} \tilde \Psi^{s_4}}{(z_4)}\, &&
\norder{\tilde \Psi^{s_5} \tilde \Psi^{s_6}}{(z_5)}
\Big\rangle\, .
\end{aligned}
\end{multline}
Their sum gives a non-vanishing parity-odd contribution to the effective
action:
\begin{equation}
\label{e:oddeven}
{\cal L}^{1+2}_{\text{odd/even}} = (-\tfrac{3}{2}+1)\, \epsilon^{mr_1\cdots r_9} t_8^{s_1\cdots s_8}
W_{r_1r_2s_1s_2}W_{r_3r_4s_3s_4}W_{r_5r_6s_5s_6}
\,H_{r_7r_8 m} H_{s_7s_8r_9}\, ,
\end{equation}
which cannot be cancelled by subtraction of tree-level graphs formed
from lower-order vertices.  Fortunately, there is a third odd/even
amplitude,
\begin{multline}
{\cal A}^3_{\text{odd/even}} = \tfrac{1}{2}
k^{(1)}_{[r_7} B^{(1)}_{r_8 m]} 
k^{(2)}_{s_7} B^{(2)}_{r_9 s_8} \,
W^{(3)}_{r_1r_2 s_1 s_2}
W^{(4)}_{r_3r_4 s_3 s_4}
W^{(5)}_{r_5r_6 s_5 s_6}\\[1ex]
\begin{aligned}
\times \Big\langle 
\norder{\Psi^{r_7} \Psi^{r_8} \Psi^{m}}{(y)}\, &
\,\,\,\Psi^{r_9}(z_2)\, &&
\norder{\Psi^{r_1} \Psi^{r_2}}{(z_3)}\, &&
\norder{\Psi^{r_3} \Psi^{r_4}}{(z_4)}\, &&
\norder{\Psi^{r_5} \Psi^{r_6}}{(z_5)}\,\\
{} & \norder{\tilde \Psi^{s_7} \tilde\Psi^{s_8}}{(z_2)}\, &&
\norder{\tilde \Psi^{s_1} \tilde \Psi^{s_2}}{(z_3)}\, &&
\norder{\tilde \Psi^{s_3} \tilde \Psi^{s_4}}{(z_4)}\, &&
\norder{\tilde \Psi^{s_5} \tilde \Psi^{s_6}}{(z_5)}
\Big\rangle\, ,
\end{aligned}
\end{multline}
obtained by taking the first term in brackets in~\eqn{e:generic_corr}
and compensating for the missing world-sheet gravitino by using a
gravitino term in~\eqn{e:V00B}.\footnote{The variable~$y$ denotes the
arbitrary and non-integrated location of the gravitino
support~\eqn{e:slice}, which leaves four positions of vertex operators
to be integrated over. Translation invariance of the world-sheet
theory allows one to convert one $z_i$~into the loop parameter~$\tau$,
and integration over the spacetime centre-of-mass momentum gives a
factor of~$1/\tau_2^5$, which together with a factor of~$\tau_2^3$
from the remaining vertex operators gives the~$1/\tau_2^2$ necessary
for a modular-invariant integral. See also the comment
below~\eqn{e:V00B}.}  This term leads to an effective-action term
which precisely cancels~\eqn{e:oddeven}, thus removing the parity-odd
term altogether.\footnote{We should perhaps mention that this
cancellation mechanism is completely different from the one that
cancels the $B\wedge t_8 W^4$ term in the type-IIB action. The latter
arose because of a sign flip between the odd/even and the even/odd
sector. As a result of the GSO projection, which adds another sign
between these two terms for the type-IIA theory, this type of
cancellation mechanism can only work for one of the two type-II
theories, not for both~\cite{vafa7}.}

This mechanism could in fact have been observed much earlier, namely
in the computation of the $B\wedge t_8 W^4$ term in the effective
action. If, in this calculation, one takes the graviton vertex
operator in the $(-1,0)$~picture (instead of the $B$~operator, as is
usually done), then the amplitude comes out with a factor $3/2$ with
respect to the usual result. Inclusion of the world-sheet gravitino
terms in the $B$~vertex operator gives another contribution which
corrects this mismatch.

Finally, in the even/even spin-structure sector one encounters (apart
from covariantisation terms for the $t_8 t_8 (DH)^2 W^2$ term) the
contribution
\begin{multline}
{\cal A}_{\text{even/even}} = 
k^{(1)}_{[r_7} B^{(1)}_{r_8 n]} 
k^{(2)}_{[s_7} B^{(2)}_{m s_8]} \,
W^{(3)}_{r_1r_2 s_1 s_2}
W^{(4)}_{r_3r_4 s_3 s_4}
W^{(5)}_{r_5r_6 s_5 s_6}\\[1ex]
\begin{aligned}
\times 
\Big\langle 
\norder{\Psi^{r_7} \Psi^{r_8}}{(z_1)}\, &
\,\,\,\,\partial X^{m}(z_2)\, &&
\norder{\Psi^{r_1} \Psi^{r_2}}{(z_3)}\, &&
\norder{\Psi^{r_3} \Psi^{r_4}}{(z_4)}\, &&
\norder{\Psi^{r_5} \Psi^{r_6}}{(z_5)}\,\\
\bar\partial X^{n}(z_1)\,\,\,\, &
\norder{\tilde \Psi^{s_7} \tilde \Psi^{s_8}}{(z_2)}\,  &&
\norder{\tilde \Psi^{s_1} \tilde \Psi^{s_2}}{(z_3)}\, &&
\norder{\tilde \Psi^{s_3} \tilde \Psi^{s_4}}{(z_4)}\, &&
\norder{\tilde \Psi^{s_5} \tilde \Psi^{s_6}}{(z_5)}
\Big\rangle\, .
\end{aligned}
\end{multline}
Taking into account the contractions of the plane-wave factors (which
are not displayed above), we find that only the zero-mode term
of~\eqn{e:lrcontract} contributes. In the effective action this
amplitude therefore leads to\footnote{\label{fn:exchange}This action
  is again not necessarily the complete result in the even
  spin-structure sector, as there are five-point amplitudes from the
  $t_8 t_8 (DH)^2 W^2$ terms that have to be subtracted in order to
  obtain the full effective action. There are also still the
  five-point tree-level graviton and NS~two-form field exchange graphs
  mentioned before which have to be subtracted.}
\begin{equation}
\label{e:eveneven}
{\cal L}_{\text{even/even}} = t_8^{r_1\cdots r_8} t_8^{s_1\cdots s_8}\,
W_{r_1r_2s_1s_2} W_{r_3r_4s_3s_4}W_{r_5r_6 s_5s_6}
\, H_{r_7r_8m} H_{s_7s_8}{}^m\, .
\end{equation}

Collecting the results~\eqn{e:oddodd} and~\eqn{e:eveneven}, we finally
arrive at the following contributions to the effective action:
\begin{multline}
\label{e:finalH2R3}
{\cal L}_{H^2 W^3} = t_8^{r_1\cdots r_8} t_8^{s_1\cdots s_8}\,
W_{r_1r_2s_1s_2} W_{r_3r_4s_3s_4}W_{r_5r_6s_5s_6}
\, H_{r_7r_8 m} H_{s_7s_8}{}^m \\[1ex]
\pm \epsilon^{r_1\cdots r_9m} \epsilon^{s_1\cdots s_9}{}_m
W_{r_1r_2s_1s_2}W_{r_3r_4s_3s_4}W_{r_5r_6s_5s_6} \Big(
H_{r_7r_8s_9} H_{s_7s_8r_9} - \tfrac{1}{9}H_{r_7r_8r_9}
H_{s_7s_8s_9}\Big)\, ,
\end{multline}
the relative sign differing between the two type-II theories (we have
suppressed an unknown relative normalisation factor between the
even/even and odd/odd parts, which among other depends on the precise
coefficient of the term~\eqn{e:GSanalogue} in the effective action, as
well as on the value of the tree-level exchange graphs mentioned in
footnote~\ref{fn:exchange}).

Whether or not the $H_{\text{NS}}$ dependence of the effective action
can be fully absorbed in a modified spin connection cannot yet be
concluded on the basis of the above results (see
\dcite{Metsaev:1987zx} for a discussion of this issue based on a
sigma-model beta-function analysis). It depends, however, crucially on
there being \emph{no} terms with six open indices on the two
gauge-field strengths. This requires a subtle cancellation between the
various terms responsible for such contributions. Similarly, the
presence of such terms with six open indices would lead to the
conclusion that the effective action does \emph{not} factorise as
\begin{equation}
\label{e:wrongH2R3}
\Big(t_8^{r_1\cdots r_8} t_8^{s_1\cdots s_8} \pm 
\epsilon^{r_1\cdots r_8np} \epsilon^{s_1\cdots s_8}{}_{np}\Big)
W_{r_1r_2s_1s_2} W_{r_3r_4s_3s_4}W_{r_5r_6 s_5s_6}
\, H_{r_7r_8 m} H_{s_7s_8}{}^m \, ,
\end{equation}
in contrast to expectations expressed in the literature. 

\end{sectionunit}

\begin{sectionunit}
\title{Comparison with the linearised superfield}
\maketitle

An elegant method for deriving higher-derivative terms in the $N=1$ 
effective action is to make use of the fact that the lowest order 
on-shell supergravity theory can be expressed entirely in terms of a 
single scalar superfield, as shown by~\dcite{nils1}. Taking powers of
this superfield and integrating over sixteen fermionic coordinates
produces $W^4$~terms familiar from string-amplitude calculations
(see~\dcite{nils2} and \dcite{Kallosh:1987mb}). A similar procedure 
can be used in the type-IIB theory, where~\dcite{howe1} have shown 
that a formulation based solely on a single scalar superfield exists, 
albeit now only at the linearised level.

In the present section we would like to compare the predictions from a
scalar superfield approach with the results of the string-amplitude
calculations discussed above. As we shall see, the scalar superfield
does \emph{not} reproduce the $H^2W^3$ terms~\eqn{e:oddodd} obtained
from string theory. We will conclude the section with a discussion of
possible reasons for this discrepancy.

The expansion of the superfield in components can be found in
e.g.~\dcite{nils1}. For our purposes the terms of interest are
\begin{equation}
\label{e:superfieldtheta}
\begin{aligned}
\Delta  = \cdots + (\thbar\Gamma^{\mu \nu\rho}\theta)\, H_{\mu \nu\rho} +
 \cdots  &+ (\thbar\Gamma^{\mu \nu\kappa}\theta)(\thbar\Gamma_\kappa{}^{\rho\sigma}
  \theta) \, W_{\mu \nu\rho\sigma} + \cdots\, .
\end{aligned}
\end{equation}
(We assume the remaining gauge fields having been set to zero and
leave out the dilaton dependence as these fields are irrelevant for
our argument). Higher-order $\theta$ terms are formed from derivative
terms or from terms non-linear in the component fields; we will return 
to the latter below. The type-IIB action has
been conjectured to be given by a $\theta^{16}$ integral of some
function of the superfield (see e.g.~\dcite{Green:1998by} for more
information),
\begin{equation}
\label{e:IIBintegral}
S_{\text{IIB}} = \int\!{\rm d}^{10}x\,{\rm d}^{16}\theta\, e\,
F[\tau_0 + \Delta]\, ,
\end{equation}
where $\tau_0 = C_0 + i e^{-\phi_0}$ is a constant background. The 
integral over sixteen thetas is, in general, very hard to evaluate. 
However, already the knowledge of the superfield expansion can be
used to draw conclusions about the structure of the terms arising
from the integration; the $H^2W^3$ terms under consideration is a
case in point.

We are interested in the $H^2W^3$ terms arising from the fifth 
power of the superfield:
\begin{equation}
\label{e:Delta5}
\begin{aligned}
{\cal L}_{H^2 W^3}\Big|_{\Delta^5} 
  &= \int\!{\rm d}^{16}\theta\,
\Big((\thbar\Gamma^{m_1m_2 k}\theta)\,
  (\thbar\Gamma_k{}^{n_1n_2}\theta) W_{m_1m_2n_1n_2}\Big)^3 \,
\Big((\thbar\Gamma^{r_1r_2r_3}\theta) H_{r_1r_2r_3} \Big)^2 .
\end{aligned}
\end{equation}
This is to be compared with the expression that yields the $W^4$
terms,
\begin{equation}
\label{e:Delta4}
{\cal L}_{W^4} = \int\!{\rm d}^{16}\theta\, \Delta^4 = \int\!{\rm d}^{16}\theta\,
\Big((\thbar\Gamma^{m_1m_2 k}\theta)\,(\thbar\Gamma_{k}{}^{n_1 n_2}\theta)\,
W_{m_1m_2n_1n_2}\Big)^4\, .
\end{equation}
In order to determine the possible contractions of the two gauge field
strengths, we employ a group-theoretical argument. The $\theta^4$
expression in the second factor of~\eqn{e:Delta5} decomposes as
\begin{equation}
\label{e:sufielddecomp}
(\mathbf{16\otimes16\otimes16\otimes16})_a =
\mathbf{770} \oplus \mathbf{1050}^+\, ,
\end{equation}
on which the product of the two gauge field strengths hence is projected
(the subscript $a$ denotes the restriction to the fully antisymmetrised part).
In general, the symmetric tensor product of two third-rank antisymmetric 
tensors contains the irreducible representations
\begin{equation}
\raise2ex\hbox{$\Big(\tableau{1 1 1}\otimes\tableau{1 1 1}\Big)_s =$}
\tabstack\cdot {1}                    \oplusr
\tabstack{\tilde{\tableau{2}}}{54}     \oplusr
\tabstack{\tableau{1 1 1 1}}{210} \oplusr 
\tabstack{\tilde{\tableau{2 2}}}{770} \oplusr
\tabstack{\tilde{\tableau{2 1 1 1 1}}^+}{1050^+} \oplusr
\tabstack{\tilde{\tableau{2 1  1 1 1}}^-}{1050^-} \oplusr
\tabstack{\tilde{\tableau{2 2 2}}}{4125}\, .
\end{equation}
Explicitly, this corresponds to 
\begin{equation}
\label{e:H2expansion}
\begin{aligned}
\mathbf{210}: & \quad \tfrac{1}{3} \big(H^m{}_{r_1r_2} H_{s_1s_2m} 
      - 2 H^m{}_{r_1s_2} H_{s_1r_2m} \big)\, ,\\[1ex]
\mathbf{770}: & \quad \tfrac{2}{3} \big( H^m{}_{r_1r_2} H_{s_1s_2m} 
      + H^m{}_{r_1s_2} H_{s_1r_2m} \big) - \text{trace terms}\, ,\\[1ex]
\mathbf{1050^\pm }:& \quad 
\tfrac{1}{2} \Big[ \tfrac{1}{2}\big(H_{r_1r_2r_3}H_{s_1s_2s_3}-3
H_{r_1r_2s_3}H_{s_1s_2r_3} \big) \\[.5ex]
 & \qquad \mp \tfrac{1}{5!} \epsilon_{r_1r_2r_3s_1s_2}{}^{n_1\ldots n_5}
\cdot 5 H_{n_1n_2n_3} H_{n_4n_5s_3} \Big] - \text{trace terms}\, ,\\[1ex]
\mathbf{4125}:& \quad \tfrac{1}{2} \big( H_{r_1r_2r_3} H_{s_1s_2s_3} 
   + 3\,H_{r_1r_2s_3} H_{s_1s_2r_3} \big) - \text{trace terms}\, ,
\end{aligned}
\end{equation}
where we implicitly assume full antisymmetrisations (separately) in 
the $r$- and $s$-indices, as well as symmetry under the 
exchange $r\leftrightarrow s$.
 
Let us now compare the above result to the action~\eqn{e:finalH2R3}
obtained from string amplitudes. The~$\mathbf{1}$~and~$\mathbf{54}$
representations are proportional to the equations of motion and thus
removable by field redefinitions, so we will focus
attention on the remaining terms. For a comparison of the other
representations we have worked out the double epsilon contraction
arising from the odd/odd sector. Rather unexpectedly we find that 
\begin{equation}
\epsilon^{r_1\cdots r_9m} \epsilon^{s_1\cdots s_9}{}_m\,
W_{r_1r_2s_1s_2} W_{r_3r_4s_3s_4} W_{r_5r_6s_5s_6} \,
\Big( H_{r_7r_8r_9} H_{s_7s_8s_9} -
3\,H_{r_7r_8s_9}H_{s_7s_8r_9}\Big)=0\, .
\end{equation}
An analysis of the $H^2$ representation content shows that this identity 
combined with the more readily derived result
\begin{equation}
\epsilon^{r_1\cdots r_9m} \epsilon^{s_1\cdots s_9}{}_m\,
W_{r_1r_2s_1s_2} W_{r_3r_4s_3s_4} W_{r_5r_6s_5s_6} \,
\epsilon_{r_7r_8r_9s_7s_8}{}^{n_1\ldots n_5} H_{n_1n_2n_3} H_{n_4n_5s_9} =0 \,,
\end{equation}
together imply that the $(\epsilon^2 W^3)^{r_7r_8r_9s_7s_8s_9}$ tensor
in the string-theory effective-action term~\eqn{e:oddodd} projects to
zero the representations $\mathbf{210}$, $\mathbf{1050^-}$ and
$\mathbf{1050^+}$. The $\mathbf{4125}$ part of~\eqn{e:oddodd} is
however non-zero, in apparent contradiction with the superfield
result.

A second apparent discrepancy concerns the $t_8 t_8 (DH)^2 W^2$ terms
computed in~\dcite{Gross:1987mw} and the analogous terms with two
epsilon factors displayed in~\eqn{e:GSanalogue}. Neither of these are
produced from the scalar superfield as displayed
in~\eqn{e:superfieldtheta}.\footnote{Note that a shift $W\rightarrow
  W+D H$ in~\eqn{e:superfieldtheta} does \emph{not} resolve this
  issue, as the $DH$ term (in the $\tilde{\tableau{2 1 1}}$
  representation) is projected out by the four-fold theta
  antisymmetrisation. The presence of a~$\theta^6 D^2 H$~term in 
  the expansion of the scalar superfield might make it possible to 
  recover the $t_8 t_8 (DH)^2 W^2$ terms. However, it is unclear 
  whether such a component is compatible with the superspace Bianchi 
  identities of linearised type-IIB supergravity from which the
  expansion of $\Delta$ derives~\cite{howe1}.}  Since
the overall coefficient of~\eqn{e:oddodd} depends on the presence
of~\eqn{e:GSanalogue} these two discrepancies are related.

If it turns out that the coefficient of~\eqn{e:GSanalogue} and the
value obtained from tree-level exchange graphs are such that there is
still a need for a non-zero coefficient of the terms
in~\eqn{e:oddodd}, then we can see two sources for the failure
of~\eqn{e:Delta5} to produce the $\mathbf{4125}$ representation.
First, in a more complete superspace treatment the vielbein
determinant in the integral~\eqn{e:IIBintegral} would be replaced by
its superfield extension, the theta expansion of which might produce
additional gauge-field terms of the kind under consideration. Another
reason why one should not have expected a match is that the current
knowledge of the theta expansion of the scalar superfield is
incomplete.  While the only $W$- and $H$-dependent terms of the
superfield that have been given in the literature are the ones
displayed in~\eqn{e:superfieldtheta}, these by no means constitute the
complete story; the complete scalar superfield contains terms
non-linear in the component fields.

We expect, e.g., a term $W H \theta^6$, with some internal
contractions between the Weyl tensor and the gauge field strength, to
contribute to the $H^2 W^3$ terms that arise from the $\Delta^4$
integral. The possible terms can be classified. The product of six
fermionic coordinates has the expansion $(\otimes^6
\mathbf{16})_a=\mathbf{3696}\oplus\mathbf{4312}$.  The contractions of
one Weyl tensor and one NS-NS~three-form gauge field strength that
correspond to these two representations are $W^{mn}{}_{[r_1r_2}
H_{r_3r_4r_5]}$ and $W^{p(m}{}_{[r_1r_2} H^{n)}{}_{r_3]p}$.  Since a
$\theta^4 H^2$ expression cannot be factored out the simple
representation-theory argument given at the beginning of this section
no longer applies. It is extremely tedious to find the precise way in
which these higher-order~$\theta$~terms arise (either by solving the
Bianchi identities or by using a gauge-completion procedure), and we
therefore refrain from following that route. 

It should be clear that any conclusions drawn from the scalar
superfield which involve other gauge-field terms will similarly suffer
from our incomplete understanding of its component expansion.

\end{sectionunit}
\end{sectionunit}

\end{sectionunit}

\begin{sectionunit}
\title{Discussion and conclusions}
\maketitle

In this paper we have analysed in detail a number of string-amplitude
computations which exhibit interesting and often overlooked features. 
While motivated by an ongoing programme that aims at the full construction 
(at order $(\alpha')^3$) of higher-derivative superinvariants for the 
type-II theories, the calculations are relevant in a very general setting. 
It is important to understand these issues in detail because there are 
very few practical checks that one can make on the not yet determined
terms in the effective action.
In particular, while we were previously~\cite{Peeters:2000qj} able to 
use supersymmetry constraints to massage our string calculations, this 
is a much more computationally-intensive check once gauge fields are included.

We have focused on three main issues. The first one concerns the
application of the chiral splitting theorem. We have shown its role in
the picture-changing operation and exhibited the importance of
left/right-mixing contractions. Our second point of emphasis is the
role of world-sheet gravitino zero modes in physical vertex
operators. These lead to new terms which have so far not received much
attention, yet are extremely relevant for obtaining consistent
amplitudes (and are expected to lead to even more serious problems
if omitted at higher genus). The third feature is that of the tensor
structure arising from fermionic correlators. Here we have shown that
the effective action receives contributions from various terms
which---although they look completely different---combine in a subtle
way once the fermionic correlators are worked out and the signs of the
GSO projection are taken into account.

Finally, we have discussed the relation of our string-based approach
to the construction of higher-derivative superinvariants using
(linearised) scalar superfields. As we have shown, its structure as
well as the string amplitude calculations are as of yet not
sufficiently known to see a precise agreement between the two, and we
have suggested possible resolutions of the apparent discrepancy.

The observations made in the present letter are relevant for the full
construction of the various other terms in the superstring effective
action, including those with RR gauge fields. This project
is in progress and will be reported on in a forthcoming publication.
\end{sectionunit}

\medskip
\section*{Acknowledgements}

We thank Angelos Fotopoulos, Olaf Lechtenfeld, Kumar Narain, Bengt
Nilsson, Joe Polchinski, Ashoke Sen, Herman Verlinde and especially
Michael Green for discussions. The Mathematica package {\tt GAMMA}
developed by Ulf Gran~\cite{Gran:2001yh} has been most useful for
gamma-matrix manipulations. Similarly, the programme {\tt
  LiE}~\cite{e_cohe1} has greatly facilitated the group-theoretical
analysis in the paper.  \medskip

\noindent P.V.~acknowledges partial financial support by the European
Commission under the RTN contract HPRN-CT-2000-00148. A.W.~wishes to
acknowledge NORDITA for financial support during a large part of this 
project.

\bigskip\bigskip\bigskip
\appendix
\begin{sectionunit}
\title{Appendix}
\maketitle
\begin{sectionunit}
\title{Operator product expansions}
\maketitle
Our conventions for the normalisations of the fields are summarised by
the following operator products and the short distance behaviour of
the correlators on the torus $\mathcal{T}$ with modular parameter
$\tau$. For the bosons we use
\begin{align}
\left\langle X(z) X(w)\right\rangle_{\mathcal{T}} &\simeq -\frac{\alpha'}{2}\ln |z-w|^2 -
\frac{\pi\alpha'}{2\,\tau_2}(z-w-\bar z + \bar w)^2\, ,\\[1ex]
\label{e:expexp}
e^{i pX(z)}\, e^{i qX(w)} &=
|z-w|^{\alpha' pq}\, e^{i(p+q) X(z)} + \cdots\, ,\\[1ex]
\label{e:deldelbar}
\left\langle\partial X(z) \bar\partial X(w)\right\rangle_{\mathcal{T}} &=
\alpha'\pi\, \delta^{(2)}(z-w) - \frac{\pi\alpha'}{\tau_2}\, ,\\[1ex]
\label{e:FF}
\left\langle F(z) F(w)\right\rangle_{\mathcal{T}} &=\alpha' \delta^{(2)}(z-w) \, .
\end{align}
The fermion normalisations are fixed by
\begin{equation}
\Psi^\mu (z) \Psi^\mu (w) = \frac{\alpha'}{2}\frac{1}{z-w} + \cdots\, ,\quad
\tilde\Psi^\mu (z) \tilde\Psi^\mu (w) = \frac{\alpha'}{2}\frac{1}{\bar
  z-\bar w} + \cdots\, ,
\end{equation}
and for the ghosts one has
\begin{align}
e^{-\chi(w)} e^{\chi(z)} &= \frac{1}{z-w} + {\cal O}(1)\, ,\\[1ex]
e^{-\phi(w)} e^{\phi(z)} &= z-w + {\cal O}((z-w)^2)\, .
\end{align}
The delta function can be represented as
\begin{equation}
\label{e:deltarep}
2\pi\,\delta^{(2)}(z-w) = \bar\partial_z \frac{1}{z-w}\, .
\end{equation}
In order to check correctness of our expressions without deriving them
from the covariant formalism, one can use the following U(1) charges
inherited from the two-dimensional world-sheet complex structure,
\begin{equation}
\begin{aligned}
\Psi         &=(-\tfrac{1}{2},0)\, , &\quad 
\theta       &=(\tfrac{1}{2},0) \, , &\quad
\partial     &=(-1,0)\, ,            &\quad
\gamma       &= (\tfrac{1}{2},0)\, , \\[1ex]
\tilde\Psi   &=(0,-\tfrac{1}{2})\, , &\quad
\bar\theta   &=(0,\tfrac{1}{2}) \, , &\quad
\bar\partial &=(0,-1)\, ,            &\quad
\tilde\gamma &=(0,\tfrac{1}{2})\,
\end{aligned}
\end{equation}
together with the ones important for the details discussed in the main text,
\begin{equation}
\chi_-      =(-1,\tfrac{1}{2})\, ,\quad
\tilde\chi_+=(\tfrac{1}{2},-1)\, ,\quad
F = (-\tfrac{1}{2},-\tfrac{1}{2})\, .
\end{equation}
\end{sectionunit}

\begin{sectionunit}
\title{Gamma matrices in the helicity basis}
\maketitle
\label{a:helicity}
In the computation of the string amplitudes we make use of gamma
matrices in the so-called helicity basis, also found in appendix~A
of~\dcite{Atick:1987rs}.  The Euclidean gamma matrices with mixed
indices $(\gamma^r)_a{}^b$ in this basis are given by
\begin{equation}
\begin{aligned}
(\gamma^1)_{a}{}^{b} &=
  \frac{\sigma^1-i\sigma^2}{2}\otimes\mathbb{1}\otimes\mathbb{1}
  \otimes\mathbb{1}\otimes\mathbb{1}\, ,\\[1ex]
(\gamma^{\bar 1})_{a}{}^{b} &=
  \frac{\sigma^1+i\sigma^2}{2}\otimes\mathbb{1}\otimes\mathbb{1}
  \otimes\mathbb{1}\otimes\mathbb{1}\, ,\\[1ex]
(\gamma^2)_{a}{}^{b} &=
  \sigma^3\otimes\frac{\sigma^1-i\sigma^2}{2}\otimes\mathbb{1}
  \otimes\mathbb{1}\otimes\mathbb{1}\, ,\\[1ex]
(\gamma^{\bar 2})_{a}{}^{b} &=
  \sigma^3\otimes\frac{\sigma^1+i\sigma^2}{2}\otimes\mathbb{1}
  \otimes\mathbb{1}\otimes\mathbb{1}\, ,\\[1ex]
(\gamma^3)_{a}{}^{b} &=
  \sigma^3\otimes\sigma^3\otimes\frac{\sigma^1-i\sigma^2}{2}\otimes\mathbb{1}
  \otimes\mathbb{1}\, ,
\end{aligned}
\quad\quad
\begin{aligned}
(\gamma^{\bar 3})_{a}{}^{b} &=
  \sigma^3\otimes\sigma^3\otimes\frac{\sigma^1+i\sigma^2}{2}\otimes\mathbb{1}
  \otimes\mathbb{1}\, ,\\[1ex]
(\gamma^4)_{a}{}^{b} &=
  \sigma^3\otimes\sigma^3\otimes\sigma^3\otimes\frac{\sigma^1-i\sigma^2}{2}
  \otimes\mathbb{1}\, ,\\[1ex]
(\gamma^{\bar 4})_{a}{}^{b} &=
  \sigma^3\otimes\sigma^3\otimes\sigma^3\otimes\frac{\sigma^1+i\sigma^2}{2}
  \otimes\mathbb{1}\, ,\\[1ex]
(\gamma^5)_{a}{}^{b} &=
  \sigma^3\otimes\sigma^3\otimes\sigma^3\otimes\sigma^3\otimes\frac{\sigma^1-i\sigma^2}{2}
  \, ,\\[1ex]
(\gamma^{\bar 5})_{a}{}^{b} &=
  \sigma^3\otimes\sigma^3\otimes\sigma^3\otimes\sigma^3\otimes\frac{\sigma^1+i\sigma^2}{2}
  \, ,\\[1ex]
\end{aligned}
\end{equation}
and from here it is easy to compute products of them. The algebra
satisfied by these matrices is
\begin{equation}
 \{ \gamma^r, \gamma^{\bar s} \} = \delta^{r s}\, ,\quad
 \{ \gamma^r, \gamma^s \} = 0\, .
\end{equation}
The relation to the basis used in the main text is given by
\begin{equation}
\Gamma^{1} = (\gamma^{1} + \gamma^{\bar 1})\, ,\quad
\Gamma^{2} = -i(\gamma^{1} - \gamma^{\bar 1})\, ,
\end{equation}
and so on, which implies that
\begin{equation}
\{ \Gamma^r, \Gamma^s \}_a{}^b = 2\,\delta^{rs}(\mathbb{1})_a{}^b\, .
\end{equation}
Index raising is done by multiplying with~$\epsilon^{ab}$ on the
right; this tensor reads
\begin{equation}
\label{e:epsilondef}
\epsilon^{ab} =
-\sigma^1\otimes\sigma^2\otimes\sigma^1\otimes\sigma^2\otimes\sigma^1\, .
\end{equation}
and its inverse satisfies $\epsilon_{ab}=\epsilon^{ab}$. 
In the notation of~\dcite{Atick:1987rs} we have 
\begin{equation}
\label{e:fermiondef}
\Psi_{AS}^1 = \tfrac{1}{2}(\Psi^1 + i\Psi^2)\,,\quad
\Psi_{AS}^{\bar 1} = \tfrac{1}{2}(\Psi^1 - i\Psi^2)\, ,
\end{equation}
and so on for the other fermions. In this
basis for the gamma matrices, SO(10) spinors are represented as a five-fold
tensor product of SO(2) spinors, 
\begin{equation}
S_a = S_\pm \otimes S_\pm \otimes S_\pm \otimes S_\pm \otimes S_\pm \, .
\end{equation}
The correlators for an arbitrary number of spin fields $S_\pm $ and
complex fermions~\eqn{e:fermiondef} are known or can be derived
through bosonisation.  The ones we need can be found
in~\dcite{Atick:1987rs}.
\end{sectionunit}

\end{sectionunit}

\bibliography{izletter}
\end{document}